\documentstyle[12pt]{article}

\begin{document}
\vspace{-2.0cm}
\bigskip
\begin{center}
{\Large \bf Wigner's little group as a gauge generator in linearized gravity
theories}
\vskip 0.5cm

{\bf Tomy Scaria}\footnote{tomy@boson.bose.res.in}
and {\bf Biswajit Chakraborty}\footnote{biswajit@boson.bose.res.in}
\vskip 1.0 true cm

S. N. Bose National Centre for Basic Sciences \\
JD Block, Sector III, Salt Lake City, Calcutta -700 098, India.

\end{center}
\bigskip

\centerline{\large \bf Abstract}

We show that the translational subgroup of Wigner's little group for massless
particles in 3+1 dimensions generate gauge transformation in linearized 
Einstein gravity. Similarly a suitable representation of the 1-dimensional 
translational 
group $T(1)$ is shown to generate gauge transformation in the linearized Einstein-Chern-Simons
 theory in 2+1 dimensions. These  representations  are  derived systematically
 from  appropriate  representations of translational groups which generate
gauge transformations in gauge theories  living in  spacetime of one
higher dimension  by the technique of dimensional descent. The unified picture 
thus obtained  is compared with a similar picture available
for vector gauge theories in 3+1 and 2+1 dimensions. 
Finally, the polarization tensor of Einstein-Pauli-Fierz theory in 2+1 
dimensions is shown to split into the polarization tensors of a pair of 
Einstein-Chern-Simons theories with opposite helicities suggesting a doublet
structure for Einstein-Pauli-Fierz theory.  
\vskip 0.5cm
\begin{flushleft}
\end{flushleft}

\section{Introduction}

The concept of Wigner's little group \cite{w} is of great importance in modern 
theoretical physics. One of the best known applications of the little group
is its role in the classification of elementary particles according to their
spin and helicity quantum numbers. Also, the transformation properties of state
vectors in Hilbert space under Poincare transformation are obtained from their
transformation properties under the action of little group, using the method
of induced representation \cite{we}. More recently, the little group is
shown to have nontrivial implications  in studying the naturality of ghosts \cite{violette}.  
Apart from its various roles mentioned above,
Wigner's little group is also known to be acting as generator of gauge
transformations in various types of gauge theories \cite{we, hk} including the 
topologically massive gauge theories \cite{bc1, bcs1}. 
It was shown by Han et al. \cite{hk} that
the $E(2)$ like little group for massless particles, or more precisely the 
translational subgroup $T(2) \subset E(2)$  
generates gauge transformations for free Maxwell theory in 3+1 dimensions.   
It was  shown in \cite{bc1} that the same little group generates gauge transformations
for Kalb-Ramond theory \cite{kr} while the translational group $T(3)$ does the job in the topologically massive $B\wedge F$ theory \cite{cs}.  These theories involve 2-form gauge fields whereas Maxwell
theory is a vector gauge theory, i.e., involves 1-form gauge field. 
Now coming to lower dimensional case,
it is again proved  that  appropriate representations of translational group
$T(1)$ generate gauge transformations  for Maxwell theory and 
topologically massive Maxwell-Chern-Simons(MCS) theory in 2+1 dimensions  
\cite{bcs1}. 
It is interesting to note that all these 
representations of the little group that generate gauge transformations 
in topologically massive theories can be systematically obtained from that in one higher dimension by a method
of dimensional descent \cite{bc2}. For instance, the representations
that generate gauge transformation for MCS theories in 2+1 dimensions can be 
obtained by dimensional descent from the representation of $T(2)$ 
that generate the gauge transformation in 3+1 dimensional Maxwell theory which
involves  splitting of the representation of $T(2)$ in to a pair of $T(1)$ representations appropriate for the 
doublet of MCS theories. Interestingly, Proca theory in 2+1 dimensions, which is
not a gauge theory, can be regarded as a doublet of MCS theories \cite{bcs, bk} which are
$U(1)$ gauge theories by themselves. These $T(1)$ groups generate gauge 
transformations in each of these MCS theories and are related by complex 
conjugation. 
  Also,  the representation of $T(3)$ that generates gauge transformation
$B\wedge F$ theory in
3+1 dimensions can be similarly obtained by the same method of dimensional 
descent from the little group for
massless particles in 4+1 dimensions as shown in \cite{bc2}. (The $B \wedge F$
theory however does not form a doublet.) The 4+1 dimensional
little group generates gauge transformations in various gauge theories 
inhabiting this higher dimensional spacetime.   
All the  above mentioned theories in whose context  the role of little 
group in generating gauge 
transformations were studied are either vector or
 antisymmetric 2-form gauge fields. These studies  established that
the various representations of the little group (or more precisely, the  translational
subgroup of the little group) which generate 
gauge transformations in the theories, massless or topologically massive, are in fact related. The connection between Wigner's little
group for massless particles and gauge transformation for 3+1 dimensional 
linearized Einstein gravity, which is a 2nd rank symmetric tensor gauge theory, 
was studied in \cite{ng}. The tensor analogue of MCS theory is the
 linearized  version of the 2+1 dimensional Einstein-Chern-Simons(ECS) theory obtained 
by coupling a CS term to  pure Einstein gravity. The ECS theory is a
topologically massive gauge theory possessing a single degree of freedom
and is described a 2nd rank symmetric tensor field. 
(Unlike the Maxwell case, the bare
Einstein gravity in 2+1 dimensions is devoid of any propagating degree of freedom.) Also it was suggested in \cite{djt, d} that the 2+1 dimensional  
Einstein-Pauli-Fierz(EPF) theory 
is a doublet of a pair of ECS theories just as 2+1 dimensional Proca theory
is a doublet of MCS theories.   
 At this stage one can ask whether 
the Wigner's little group generate
gauge transformations in the
ECS theories  also. If the answer is yes, are these representations
 the same as the ones obtained in the case of vector/2-form gauge fields
and  can one employ the method
of dimensional descent in this case too? Analogous to Proca theory 
in 2+1 dimensions,
does the suggested doublet structure of EPF theory enables one to construct the 
representations that generates gauge transformations in ECS theories?
Many of  these questions are yet to be answered 
and the purpose of this
paper is to undertake a study of  these problems. Here we study the little group and dimensional descent from 3+1 dimensionsional linearized  Einstein gravity
to linearized ECS theory 
in 2+1 dimensions.
However, in the present work  our main intention is to study the role of
Wigner's little group as a generator of gauge transformation in topologically
massive ECS theory, where gauge invariance coexists with massive excitations,
like the corresponding MCS theory involving vector gauge fields.
Our approach to the problem consists of considering  the momentum space 
equation of motion using  which we obtain explicit form of the polarization
vector/tensor of the theory under consideration. Using the expression of
the  polarization vector/tensor thus obtained, we study the role of the 
respective Wigner's little group in generating gauge transformation in the 
theory. This approach, therefore, is different from the one used in \cite{ng}. 
It should be noticed that, 
on account of the fact that unlike other field theories in whose contexts
the translational group was shown to generate gauge transformations, in the 
linearized gravity theories the gauge transformation and spacetime 
diffeomorphisms are 
intertwined.  Thus the problem is interesting in its own right.

The paper is organized as follows. Section 2 provides a brief review of the role of 
little group as a gauge generator in the usual and topologically massive gauge 
theories. In the context  of topologically massive gauge theories like the MCS and $B\wedge F$ theories, we provide a brief outline of    the method of dimensional descent.
In section 3, we study the 
role of little group as a generator of gauge transformation in linearized 
Einstein gravity in 3+1 dimensions and and in 2+1 dimensional ECS theory.
Section 4 discusses the doublet structure of EPF theory in 2+1 dimensions
and applies the method of dimensional descent from 3+1 dimensional 
linearized gravity to 2+1 dimensional EPF theory. We also explain in section 4,
how the representations of $T(1)$ that generates gauge transformation in the 
doublet of ECS theories can be obtained by a careful study of the
polarization tensors of EPF and ECS theories. We finally conclude in section 5.

\section{Wigner's little group and dimensional descent in vector and 2-form
gauge theories}
Wigner's little group is defined as the subgroup of Lorentz group that
preserves the energy-momentum vector of a particle. We begin by reviewing the 
role of Wigner's little group as a
generator of gauge transformation in Maxwell  and Kalb-Ramond (KR) theories 
in 3+1 dimensions and also of the method of dimensional descent. 
Our notation is as follows. We use   $\mu, \nu $ etc to denote the spacetime
indices in 3+1 dimensions and $a, b$ etc for  2+1 dimensions. Signature of the
metric is `mostly negative'.
\subsection{Wigner's little group and gauge transformation  in 
Maxwell and KR theories} 

An element of the little group that
leaves the four-momentum $k^\mu = (\omega, 0, 0, \omega )^T$ of a massless 
particle moving in the $z$-direction invariant is given by \cite{hk}
\begin{equation}
W_4(p, q;  \phi) =
\left( \begin{array}{cccc}
1+ \frac{p^2 + q^2 }{2} & p\cos \phi - q \sin \phi & q\sin \phi + p \cos \phi  & -\frac{p^2 + q^2 }{2} \\
p & \cos \phi & \sin \phi  & -p \\
q & -\sin \phi & \cos \phi  & -q \\
\frac{p^2 + q^2 }{2} &  p\cos \phi -q \sin \phi & q\sin \phi + p \cos \phi  & 1 -\frac{p^2 + q^2 }{2}
\end{array}\right).
\label{1}
\end{equation}
Here $p,q$ are  any real numbers. 
This little group can be written as 
\begin{equation}
W_4(p, q;  \phi) = W(p, q) R(\phi)
\label{2}
\end{equation}
where 
\begin{equation}
W(p, q) = W_4(p, q;  0) = \left( \begin{array}{cccc}
1+ \frac{p^2 + q^2 }{2} & p & q  & -\frac{p^2 + q^2 }{2} \\
p & 1 & 0 & -p \\
q & 0 & 1 & -q \\
\frac{p^2 + q^2 }{2} &  p & q  & 1 -\frac{p^2 + q^2 }{2}
\end{array}\right)
\label{3}
\end{equation}
is  a particular representation of the  translational subgroup of $T(2)$ of
the little group. We first discuss how this representation of $T(2)$ 
generates gauge transformation in Maxwell theory.
The Lagrangian for Maxwell theory is 
\begin{equation}
{\cal{L}} = -\frac{1}{4}F_{\mu \nu }F^{\mu \nu }, ~~~~~F^{\mu \nu } = 
\partial_\mu A_\nu - \partial_\nu A_\mu 
\label{4} 
\end{equation}
and the corresponding equation of motion is \begin{equation}
\partial_\mu F^{\mu \nu} = 0
\label{5}
\end{equation}
For a single mode, without loss of generality,
  the gauge field $A^\mu (x)$ can be written as
\begin{equation}
A^{\mu}(x) = \varepsilon^{\mu}(k) e^{ik \cdot x}
\label{6}
\end{equation}
suppressing the positive frequency part for simplicity. Here $\varepsilon^{\mu}$ 
stands for the polarization vector of the photon\footnote{This method \cite{felsager}, henceforth referred to as ``plane wave method", 
 was used in \cite{bc1, bcs} to find  masses of the excitations and  structures of  the maximally
reduced form of the  polarization  
vectors/tensors in various theories. }. 
In terms of the 
polarization vector, the gauge transformation $A_{\mu}(x) \rightarrow 
A^{\prime}_{\mu} =  A_{\mu} + \partial_{\mu}f$ (where $f(x)$ is an arbitrary 
scalar function) is expressed  as
\begin{equation}
\varepsilon_{\mu}(k) \rightarrow \varepsilon^{\prime}_{\mu} =  \varepsilon_{\mu}(k) + if(k)k_{\mu}
\label{7}
\end{equation}
where $f(x)$ has been written as $f(x) = f(k)e^{ik \cdot x}$. 
The equation of motion, in terms of the polarization vector,  will now be given by
\begin{equation}
k^2 \varepsilon^{\mu} - k^{\mu} k_{\nu} \varepsilon^{\nu} = 0
\label{8}. 
\end{equation}
The massive excitations corresponding to $k^2 \neq 0$ leads to the solution
$\varepsilon^{\mu} \propto k^{\mu}$ which can therefore be gauged away. For
massless excitations  ($k^2 = 0$), the Lorentz condition $k_\mu \varepsilon^{\mu} =0$ follows immediately from (\ref{8}).  Taking  $k^{\mu} = (\omega, 0, 0, \omega)^T$, corresponding to a photon of energy $\omega$ propagating in the $z$
direction, one can easily show that $\varepsilon^{\mu}(k)$ takes the form
\begin{equation}
\varepsilon^{\mu}(k) =   
(0, \varepsilon^1, \varepsilon^2, 0)^T 
\label{9}
\end{equation}
up to a gauge transformation. Note that (\ref{9}) is the maximally reduced form 
of $\varepsilon^{\mu}$ displaying the two transverse degrees of freedom
$\varepsilon^1$ and $\varepsilon^2$. 
Under the action (\ref{3}) of the translational group $T(2)$  this polarization 
vector transforms  as follows:
\begin{equation}
\varepsilon^{\mu} \rightarrow \varepsilon^{\prime \mu} = {W^{\mu}}_{\nu}(p, q) \varepsilon^{\nu} = \varepsilon^{\mu} +
 \left( \frac{p\varepsilon^1 + q\varepsilon^2}{\omega}\right)k^{\mu}~.
\label{10}
\end{equation}
Clearly, using (\ref{7}), this can be identified as a gauge transformation by
choosing $f(k)$ suitably. 

By similar methods it can shown that gauge
transformations are generated by $T(2)$ in the 3+1 dimensional 
Kalb-Ramond(KR) theory \cite{bc1} described by the  Lagrangian
\begin{equation}
{\cal{L}} = \frac{1}{12}H_{\mu \nu \lambda}H^{\mu \nu \lambda}
\label{11}
\end{equation}
where the 3rd rank field strength tensor $H_{\mu \nu \lambda}$ is given by
\begin{equation}
H_{\mu \nu \lambda} = \partial_{\mu}B_{\nu \lambda} + \partial_{\nu} B_{\lambda
\mu} + \partial_{\lambda} B_{\mu \nu}
\label{12}
\end{equation}
with $B_{\mu \nu} = -B_{\nu \mu}$ being the 
rank-2 antisymmetric gauge field. 
The equation of motion for the KR theory is given by
\begin{equation}
\partial_\mu H^{\mu \nu \lambda} = 0.
\label{13}
\end{equation}
Here the model is invariant under the reducible gauge transformation given by\footnote{This is reducible, since by the  choice  $f_{\mu} = \partial_{\mu}\Lambda$, of the arbitrary functions $f_{\mu}$, the gauge variation can be made to vanish.}
\begin{equation}
B_{\mu \nu} \rightarrow B^{\prime}_{\mu \nu} = B_{\mu \nu} + \partial_\mu f_\nu
- \partial_\nu f_\mu ~.
\label{14}
\end{equation}
Analogous to Maxwell case, we follow the plane wave method and adopt for KR theory, the ansatz 
\begin{equation}
B^{\mu \nu}(x) = \varepsilon^{\mu \nu}(k)e^{i k \cdot x}
\label{15}
\end{equation}  
where $\varepsilon^{\mu \nu}(k)$ is the antisymmetric polarization tensor of
KR theory. In terms of the polarization tensor, the equation of motion (\ref{13})
can be written as 
\begin{equation}
k_{\mu}[k^{\mu} \varepsilon^{\nu \lambda} + k^{\nu} \varepsilon^{\lambda \mu} +
k^{\lambda} \varepsilon^{\mu \nu}] =0
\label{16}
\end{equation} 
and the gauge transformation (\ref{14}) as
\begin{equation}
\varepsilon_{\mu \nu} \rightarrow \varepsilon^{\prime}_{\mu \nu} = \varepsilon_{\mu \nu} + i(k_{\mu}f_{\nu}(k) - k_{\nu}f_{\mu}(k)).
\label{17}
\end{equation}
For the case $k^2 \ne 0$, from (\ref{16}) one has 
\begin{equation}
\varepsilon^{\nu \lambda} = \frac{1}{k^2}[k^{\nu}(k_{\nu}\varepsilon^{\mu \lambda}) - k^{\lambda}(k_{\mu}\varepsilon^{\mu \nu})].
\label{18}
\end{equation}
These (massive) solutions can be gauged away  by choosing
$f^{\lambda}(k) = \frac{i}{k^2}k_{\mu}\varepsilon^{\mu \lambda}$. Similar  to
Maxwell case, here too the massive excitations  are gauge artifacts.
As for $k^2 = 0$, we have from (\ref{16}) 
$ k_{\mu}\varepsilon^{\mu \nu} = 0$ which is the similar to the Lorentz
condition in Maxwell theory. Using this condition and by choosing a frame
where $k^\mu = (\omega, 0, 0, \omega)^T$ just as in Maxwell theory, the polarization tensor reduces to
the maximally reduced form
\begin{equation}
{\cal E} \equiv \{\varepsilon^{\mu \nu} \} = \varepsilon^{12}
  \left( \begin{array}{cccc}
0 & 0 & 0 & 0 \\
0 & 0 & 1 & 0 \\
0 & -1 & 0 & 0 \\
0 & 0 & 0 & 0
 \end{array} \right)
\label{19}
\end{equation}
after a suitable gauge transformation that does away with the spurious degrees of
freedom. In the maximally reduced form of ${\cal E}$ (\ref{19}) one is
left with the only physical  degree of freedom corresponding to $\varepsilon^{12} $ \cite{bc1}. 
Since under Lorentz transformation $\varepsilon^{\mu \nu}$ transforms as
  $\varepsilon^{\mu \nu} \rightarrow \varepsilon^{\prime \mu \nu} = 
{\Lambda^{\mu}}_\rho {\Lambda^{\nu}}_\sigma \varepsilon^{\rho \sigma}$ 
the transformation of ${\cal E}$ matrix under the action of the little
group $W(p, q)$, which in this case is subgroup of Lorentz group, is given by
\begin{equation}
{\cal E} \rightarrow {\cal E} ^{\prime} = W(p,q) {\cal E} W^T (p,q) \\
= \varepsilon^{12}\left( \begin{array}{cccc}
0 & -q & p & 0 \\
q & 0 & 1 & q \\
-p & -1 & 0 & -p \\
0 & -q & p & 0
\end{array} \right) = {\cal E} + \varepsilon^{12}\left( \begin{array}{cccc}
0 & -q & p & 0 \\
q & 0& 0 & q \\
-p & 0 & 0 &  -p \\
0 & -q & p & 0
\end{array} \right).
\label{20}
\end{equation}
With the choice $f^1 = \frac{iq}{\omega}\varepsilon^{12},
f^2 = - \frac{ip}{\omega}\varepsilon^{12} $
and $f^3 = f^0$, one can write (\ref{20}) as a gauge transformation (\ref{17}).
Thus, the translational group $W(p,q)$ generates gauge transformation in KR theory
also.

By similar methods it was shown that the translational group $T(3)$ generates
gauge transformation in the topologically massive $B\wedge F $ theory which has the Lagrangian
\begin{equation}
{\cal L} = -\frac{1}{4}F_{\mu \nu}F^{\mu \nu} + \frac{1}{12}H_{\mu \nu \lambda}H^{\mu \nu \lambda} - \frac{m}{6}\epsilon^{\mu \nu \lambda \rho} H_{\mu \nu \lambda} A_{\rho}, ~~~~~~m > 0
\label{21}
\end{equation}
which is obtained by topologically coupling the $B_{\mu \nu}$ field of
 Kalb-Ramond theory(\ref{11}) with the Maxwell field $A_{\mu}$ so that the last term in
(\ref{21}) does not contribute to the  energy-momentum tensor.
 The  B$\wedge$F Lagrangian (\ref{21}) can be regarded 
either as a massive
Maxwell(i.e., Proca) theory or a massive KR theory \cite{bw}. 
For the sake of completeness, we summarize here the essential points
concerning the role of $T(3)$ in generating gauge transformation in $B\wedge F$
theory \cite{bc1}. Following the plane wave method, the polarization vector 
and tensor for this theory, in the  
maximally reduced form (analogous to (\ref{9}) and (\ref{19})), are found to be
\begin{equation}
\varepsilon^{\mu} = -i \left( \begin{array}{c}
0 \\
a \\
b \\
c
\end{array} \right) ~~~~~~~~
{\cal E} = \left( \begin{array}{cccc}
0 & 0 & 0 & 0 \\
0 & 0 & c & -b \\
0 & -c & 0 & a \\
0 & b & -a & 0
\end{array} \right)
  \label{22}
\end{equation}
where $ a, b, c$ being three arbitrary real parameters corresponding to three 
degrees of freedom and we have chosen the rest frame
where 
\begin{equation} 
k^{\mu} = (m, 0, 0, 0)^T.
\label{22+1}
\end{equation}
Note that the number of degrees of freedom for B$\wedge$F theory can be obtained by adding 
the number of degrees of freedom for Maxwell and KR theories together 
\cite{bc1}. Clearly, unlike the Maxwell or KR theories,
the 2-dimensional  translational group $T(2)$ cannot generate gauge transformation in
$B\wedge F$ theory having three degrees of freedom. It is the 3-dimensional translational group $T(3)$ 
represented by
\begin{equation}
D(p, q, r) = 1+ pT_1 + qT_2 + rT_3 = \left( \begin{array}{cccc}
1 & p & q & r  \\
0 & 1 & 0 & 0 \\
0 & 0 & 1 & 0 \\
0 & 0 & 0 & 1
\end{array} \right)
\label{23}
\end{equation}
that generate gauge transformation in this case. Here  $T_1 = \frac{\partial D}{\partial p}, T_2 = \frac{\partial D}{\partial q},T_2 = \frac{\partial D}{\partial r}$ are the Lie algebra generators of the abelian algebra of $T(3)$. One can immediately 
see that 
\begin{equation}
\delta \varepsilon^{\mu} = {D^{\mu}}_{\nu}(p, q, r) \varepsilon^{\nu} - \varepsilon^{\mu} = \frac{i}{m} (pa+qb+rc) k^{\mu} 
= (pT_1 + qT_2 + rT_3) \varepsilon^{\mu}
\label{24}
\end{equation}
and 
\begin{equation}
\delta {\cal E} = D(p, q, r) {\cal E} D^T(p, q, r) - {\cal E} = \left( \begin{array}{cccc}
0 & (r b- q c) & (p c - r a) & (q a - p b) \\
-(r b- q c) & 0 & 0 & 0 \\
-(p c - r a) & 0 & 0 & 0 \\
-(q a - p b) & 0 & 0 & 0
\end{array} \right)
\label{25}
\end{equation}
so that using (\ref{7}) and (\ref{17}) one can cast (\ref{24}) and (\ref{25})
respectively in 
the form of gauge transformations.  Further it has been shown in 
\cite{bc2} that the representation (\ref{23}) of $T(3) \subset E(3)$ can be constructed from one higher i.e., 4+1 dimensional spacetime by a method of 
`dimensional descent' which involves  the projection operator ${\cal P}$ = diag$(1, 1, 1,1, 0
)$. This is expected, as $ E(3)$ is the Wigner's little  group for massless
particles in 4+1 dimensions and under this projection, the polarization vector 
(tensor) of free Maxwell (KR) theory and the momentum 5-vector for a massless photon moving in the 4th direction 
in 4+1 dimensional spacetime maps to the polarization vector $\varepsilon^\mu$ 
(tensor  $\varepsilon^{\mu \nu}$)
(\ref{22}) and momentum 4-vector $k^\mu$ (\ref{22+1}) of a massive B$\wedge$F
quanta at rest in 3+1 dimensions. The same feature of ``dimensional descent"
goes through from 3+1 dimensions to 2+1  dimensions so that starting from 
usual photon in 3+1 dimensions and using  ${\cal P}$ = diag$(1, 1, 1, 0)$
one obtain the polarization vector and momentum 3-vector of a Proca quanta 
governed by the Proca theory  
\begin{equation}
{\cal L} = -{\frac{1}{4}}{F^{ab}F_{ab}} + {\frac{\omega^2}{2}}A^{a}A_{a}
\label{26}
\end{equation}
in 2+1 dimensions. In order to discuss the gauge transformation properties
it is essential to provide a $3 \times 3$ representation of $T(2)$ (denoted by
$\bar{D}(p, q)$) which now amounts  to deleting the last row and column of $D(p, q, r)$ in (\ref{23}),
\begin{equation}
\bar{D}(p, q) = \left(
\begin{array}{ccc}
1 & p & q   \\
0 & 1 & 0   \\
0 & 0 & 1   \\
\end{array}
\right).
\label{27}
\end{equation}
The corresponding generators are given by,
\begin{equation}
\bar{T}_1 = \frac{\partial \bar{D}}{\partial p} = \left(
\begin{array}{ccc}
0 & 1 & 0 \\
0 & 0 & 0 \\
0 & 0 & 0
\end{array}
\right); \hskip 1.0cm \bar{T}_2 = \frac{\partial \bar{D}}{\partial q} =
\left( \begin{array}{ccc}
0 & 0 & 1 \\
0 & 0 & 0 \\
0 & 0 & 0
\end{array}
\right)
\label{28}
\end{equation}

Just as the Proca theory in 3+1 dimensions maps to the $B \wedge F $ theory,
where gauge transformations  were discussed, the Proca theory in 2+1 dimensions
is actually a doublet of
Maxwell-Chern-Simons theories,\cite{bcs, bw, d}
\begin{equation}
{\cal L} = {\cal L}_+ \oplus {\cal L}_-
\label{29}
\end{equation}
where
\begin{equation}
{\cal L}_\pm = -{\frac{1}{4}}{F^{ab}F_{ab}} \pm {\frac{{\theta
}}{2}}{\epsilon}^{abc}A_{a}{\partial}_{b}A_{c}
\label{30}
\end{equation}
with each of ${\cal L}_+$ or ${\cal L}_-$ being a topologically massive gauge
 theory. The mass of the MCS quanta is $\omega = |\theta|$, where $\omega$
 is the parameter entering in (\ref{26}). We can therefore study the gauge transformation generated in this
doublet.
The polarization vector for
${\cal L}_\pm $, with only one degree of freedom for each of ${\cal L}_+$ and
${\cal L}_-$, has been found to be\cite{bcs, grignani}
\begin{equation}
\bar{\varepsilon}^{a}_\pm = \frac{1}{\sqrt 2} \left( \begin{array}{c}
0 \\
1 \\
\pm i
\end{array}
\right)
\label{31}
\end{equation}
while the 3-momentum $k^a = (\omega , 0, 0)^T$ obviously takes the same form as in the Proca model.
In analogy with (\ref{24}), here also one can write,
\begin{equation}\delta \bar{\varepsilon}^{a} = {(p\bar{T}_1 + q\bar{T}_2)^a}_b \bar{\varepsilon}^{b}
\label{32}
\end{equation}
where $\bar{\varepsilon}^{a} = (0, a_1, a_2 )^T$
 is the polarization vector for the Proca theory in 2+1 dimensions.
Had the Proca theory been a gauge theory, (\ref{32}) would
have represented a gauge transformation, as it can be written as
\begin{equation}
\delta \bar{\varepsilon}^{a} = \frac{pa_1 + qa_2}{\omega}k^a~.
\label{33}
\end{equation}
But since Proca theory is not a gauge theory, we can only study the gauge transformation properties of each of the doublet ${\cal L}_\pm$ (\ref{30})
individually. First note that the
 Proca polarization vector $\bar{\varepsilon}^{a}$ is just a linear
combination of the two real orthonormal canonical vectors $\varepsilon_1$
and $ \varepsilon_2$ where,
\begin{equation}
\bar{\varepsilon}^{a} = a_1 \varepsilon_1 + a_2 \varepsilon_2; \hskip 1.0 cm
\varepsilon_1 = (0,1,0)^T,
\varepsilon_2 = (0,0,1)^T.
\label{34}
\end{equation}
Correspondingly the generators $\bar{T}_1$ and $\bar{T}_2$  (\ref{28}),
form an
orthonormal basis as they satisfy  $tr(\bar{T}_a^{\dagger}\bar{T}_b) = \delta_{ab}$. Furthermore,
\begin{equation}
\bar{T}_1\varepsilon_1 = \bar{T}_2\varepsilon_2 = (1, 0, 0)^T
= \frac{k^a}{\omega}, \hskip 1.0 cm \bar{T}_1\varepsilon_2 = \bar{T}_2\varepsilon_1 = 0
\label{35}
\end{equation}
On the other hand, the  polarization vectors
$\bar{\varepsilon}^{a}_+$ and $\bar{\varepsilon}^{a}_-$ (\ref{31})
also provide an orthonormal basis(complex) in the plane as
\begin{equation}
(\bar{\varepsilon}^{a}_+)^{\dagger}(\bar{\varepsilon}^{a}_-) = 0;
\hskip 1.0cm (\bar{\varepsilon}^{a}_+)^{\dagger}(\bar{\varepsilon}^{a}_+) =
(\bar{\varepsilon}^{a}_-)^{\dagger}(\bar{\varepsilon}^{a}_-) = 1.
\label{36}
\end{equation}
Here we note that spatial part $\vec{\varepsilon}_\pm$ of ${\varepsilon}_\pm$ 
 can be obtained from the space part of the  above mentioned canonical ones by appropriate
$SU(2)$ transformation. That is, $U= \frac{1}{\sqrt{2}}\left( \begin{array}{cc} 
1 & i \\
i & 1 \end{array} \right) 
\in $ SU(2) when acts on $\vec{\varepsilon}_1 = \left( \begin{array}{c} 1 \\ 0 \end{array} \right)$ 
and $\vec{\varepsilon}_2 = \left( \begin{array}{c} 0 \\ 1 \end{array} \right)$ 
yields respectively the vectors $\vec{\varepsilon}_+ = \frac{1}{\sqrt{2}}\left( \begin{array}{c} 1 \\ i \end{array} \right)$ and  $\vec{\varepsilon}_- = \frac{1}{\sqrt{2}}\left(
\begin{array}{c} 1 \\ -i \end{array} \right)$
(up to an irrelevant factor of $i$)\footnote{This ambiguity of $i$ factor is 
related to the U(1) phase arbitrariness of the polarization vector \cite{bcs}.}:
\begin{equation} 
\vec{\varepsilon}_+ = U \vec{\varepsilon}_1, ~~~~\vec{\varepsilon}_- = iU \vec{\varepsilon}_2.
\label{37-1}
\end{equation} 
 This suggests that we consider the following
 orthonormal basis for the Lie algebra of $T(2)$:
\begin{equation}
\bar{T}_\pm = \frac{1}{\sqrt{2}}(\bar{T}_1 \mp i\bar{T}_2) = \frac{1}{\sqrt{2}}\left( \begin{array}{ccc}
0 & 1 & \mp i \\
0 & 0 & 0 \\
0 & 0 & 0 \end{array} \right)
\label{37}
\end{equation}
instead of $\bar{T}_1$ and $\bar{T}_2$. Note that they also satisfy relations similar to
the (1-2) basis,
\begin{equation}
tr(\bar{T}_+^{\dagger}\bar{T}_+) = tr(\bar{T}_-^{\dagger}\bar{T}_-) = 1; tr(\bar{T}_+^{\dagger}\bar{T}_-) = 0
\label{38}
\end{equation}
One can now easily see that
\begin{equation}
\bar{T}_+{\varepsilon}_+ = \bar{T}_-\varepsilon_- = \frac{k^a}{\omega}, \hskip 1.0cm \bar{T}_+{\varepsilon}_- = \bar{T}_-\varepsilon_+ = 0
\label{39}
\end{equation}
analogous to (\ref{35}). Furthermore,
\begin{equation}
\delta \bar{\varepsilon}^{a}_\pm = \alpha_{\pm}\bar{T}_{\pm} \bar{\varepsilon}^{a}_\pm =
 \frac{\alpha_{\pm} }{\omega}k^a.
\label{40}
\end{equation}
This indicates that $\bar{T}_\pm$ - the generators of the Lie algebra of $T(2)$
in the
rotated (complex)basis -
generate independent  gauge transformations in ${\cal L}_{\pm}$ respectively. One
therefore can understand how the appropriate representation
 of the generator of gauge
transformation in the doublet of MCS theory can be obtained from higher
3+1 dimensional Wigner's group through dimensional descent.
A finite gauge transformation is obtained by integrating (\ref{40}) i.e., exponentiating the corresponding Lie algebra element. This gives two
representations of Wigner's little group for massless particles in $2+1$
dimensions, which is isomorphic to ${\cal R}\times {\cal Z}_2$, although here 
we are just considering the component which is connected to the identity,
\begin{equation}
G_\pm (\alpha_\pm) = e^{\alpha_\pm \bar{T}_\pm} = 1 + \alpha_\pm \bar{T}_\pm = \left(
\begin{array}{ccc}
1 & \frac{\alpha_\pm}{\sqrt{2}} & \mp i \frac{\alpha_\pm}{\sqrt{2}} \\
0 & 1 & 0 \\
0 & 0 & 1
\end{array}
\right)
\label{41}
\end{equation}
Note that $G_\pm (\alpha_\pm) $ generates gauge transformation in the
        doublet ${\cal L}_{\pm}$,  
\begin{equation}
{G_\pm^a}_b \bar{\varepsilon}^{b}_\pm =  \bar{\varepsilon}^{a}_\pm + 
\frac{\alpha_\pm}{|\mu |} k^a 
\label{42}
\end{equation}
and are related by complex conjugation.
This
complex conjugation is also a symmetry of the doublet \cite{bcs}.

Therefore, it is clear that suitable representations of  translational groups
in different dimensions  act as generators of gauge transformations in 
 ordinary as well as topologically massive gauge theories. We now turn our
attention to the case of linearized gravity.

\section{Role of little group in linearized gravity theories}
 
Gravity (linearized) in $d$ spacetime dimensions is described by a symmetric  second rank tensor gauge field and  has $\frac{1}{2}d(d-3)$ degrees of 
freedom\footnote{The degree of freedom counting can be done by following 
Weinberg \cite{weinberg1}. 
To start with, note that a symmetric second rank tensor in $d$ dimensions has 
$\frac{1}{2}d(d+1)$ independent components. Analogous to 
the Lorentz gauge condition ($\partial^\mu A_\mu = 0$) of Maxwell theory, 
in general relativity we have the harmonic gauge condition 
$g^{\mu \nu} {\Gamma^\lambda}_{\mu \nu} = 0$ which amounts to $d$ constraints 
on the components of
$g_{\mu \nu}$. These along with the $d$ independent components of the   gauge parameter (which by itself is a $d$-vector now;
see the ensuing discussion below, particularly (\ref{47})), in the linearized 
version of the theory, reduces the number of independent components of the 
tensor 
field to $ \frac{1}{2}d(d+1) - 2d = \frac{1}{2}d(d-3)$.}. Therefore general 
relativity in 3+1 dimensions has two degrees of freedom and in 2+1 dimensions 
it has none. However, 2+1 dimensional gravity coupled to a Chern-Simons
 topological term, with gauge group being the Lorentz group itself, 
possesses a single propagating massive degree of freedom \cite{djt}. Just like the MCS theory,
the gauge invariance coexists with mass in the linearized version of this theory too where the gauge group reduces to abelian group $T(1)$. In this section we
study the role of  translational group in generating gauge transformations
in the linearized versions of  Einstein theory in 3+1 dimensions and gravity 
coupled to Chern-Simons term 2+1 dimensions. 

Following the same conventions as of \cite{djt}, we write the pure
Einstein action in 3+1 dimensions as
\begin{equation}
I^E = -\int d^4 x  {\cal L}^E,  ~~~~~ {\cal L}^E = \sqrt{g} R = \sqrt{g} g^{\mu\nu}R_{\mu\nu}
\label{43}
\end{equation} 
where ${\cal L}^E$ is the Einstein Lagrangian and $R_{\mu\nu}$ is the Ricci 
tensor. 
  In the linearized approximation the metric $g_{\mu\nu}$ is 
assumed to be close to the  flat background part $\eta^{\mu\nu}$ and therefore\footnote{We have set the parameter $\kappa$ appearing in \cite{djt} equal to 
unity.}
\begin{equation} 
g_{\mu\nu} = \eta_{\mu\nu} + h_{\mu\nu}
\label{43+1}.    
\end{equation}
where $h_{\mu\nu}$ is the deviation such that $|h_{\mu\nu}|<< 1 $. When the 
deviation is small one considers only terms up to first order
in $h_{\mu\nu}$. The raising and lowering of indices is done using $\eta_{\mu\nu}$.
\subsection{Linearized Einstein gravity in 3+1 dimensions}
The linearized version of 
Einstein-Hilbert Lagrangian  is
\begin{equation}
{\cal L}_L^E = \frac{1}{2}h_{\mu\nu} \left[ R^{\mu\nu}_L - \frac{1}{2} \eta^{\mu\nu} R_L\right]. 
\label{44}
\end{equation} 
Here $ R^{\mu\nu}_L$ is the linearized Ricci tensor given by
\begin{equation}
 R^{\mu\nu}_L = \frac{1}{2}(- \Box h^{\mu\nu} + \partial^\mu \partial_\alpha  h^{\alpha \nu} + \partial^\nu\partial_\alpha  h^{\alpha \mu} - \partial^\mu \partial^\nu h)
\label{45}
\end{equation}
with $h = h^\alpha_\alpha$. Similarly $R_L = {R^\alpha_L}_\alpha$. The field
equations for $ h^{\mu\nu}$ following from this Lagrangian is given by
\begin{equation}
- \Box h^{\mu\nu} + \partial^\mu \partial_\alpha  h^{\alpha \nu} + \partial^\nu\partial_\alpha  h^{\alpha \mu}
 - \partial^\mu \partial^\nu h + \eta^{\mu\nu}(\Box h -\partial_\alpha \partial_\beta h^{\alpha \beta}) = 0.
\label{46}
\end{equation}
The above equation is invariant under the following gauge transformation:
\begin{equation}
h_{\mu\nu} \rightarrow h'_{\mu\nu} = h_{\mu\nu} + \partial_\mu \zeta_\nu (x) + \partial_\nu  \zeta_\mu (x) 
\label{47}
\end{equation}
which is just the symmetric counterpart of the gauge transformation (\ref{14}) 
for KR and $B\wedge F$ theories. However, unlike KR and $B\wedge F$ theories, 
linearized gravity is not a reducible gauge system.  
Here $ \zeta_\mu(x)$ are completely arbitrary except that they
 are considered to be small. 
Following the plane wave method,  we now adopt  the ansatz
\begin{equation}
h_{\mu\nu} = \varepsilon_{\mu\nu} (k) e^{ik.x} + c.c.
\label{48}
\end{equation}
where $\varepsilon_{\mu\nu}$ is the symmetric  polarization tensor and is the 
counterpart of 
the antisymmetric polarization tensor appearing in (\ref{15}) for KR theory.
With the choice 
\begin{equation}
 \zeta_\mu(x) = -i\zeta_\mu(k) e^{ik.x} + c.c.
\label{49}
\end{equation}
the gauge transformation in $h_{\mu\nu}$ can be written in terms of the 
polarization tensor as 
\begin{equation} 
 \varepsilon_{\mu\nu} (k)  \rightarrow {\varepsilon}'_{\mu\nu} (k) = \varepsilon_{\mu\nu} (k) + k_\mu \zeta_\nu (k) +  k_\nu \zeta_\mu (k).
\label{50}
\end{equation}
Just as in the Maxwell case, hereafter we  will consider only the the negative 
frequency  part for simplicity.  
Substituting the ansatz in the equation of motion yields
\begin{equation}
k^2\varepsilon^{\mu\nu} - k^\mu k_\alpha \varepsilon^{\alpha \nu} - 
k^\nu k_\alpha \varepsilon^{\alpha \mu }
+  k^\mu k^\nu \varepsilon   
+\eta^{\mu\nu}(-k^2 \varepsilon +k_\alpha k_\beta \varepsilon^{\alpha \beta}) = 0.
\label{51}
\end{equation}

As was done  in the previous cases, we separately consider the two possibilities $k^2 \ne 0$ and $k^2 = 0$. Choosing the massive ( $k^2 \ne 0$) case first, 
we contract the equation of motion (\ref{51}) with $\eta^{\mu\nu}$ to obtain
\begin{equation}
k^2\varepsilon - k_\alpha k_\beta \varepsilon^{\alpha \beta} = 0~;~~~~ \varepsilon = {\varepsilon^\mu}_\mu.
\label{51+1}
\end{equation}   
A general solution to this equation is given by
\begin{equation}
\varepsilon^{\mu\nu} = k^{\mu} f^{\nu}(k) +  k^{\nu} f^{\mu}(k)
\label{51+2}
\end{equation}
with $f^{\nu}(k)$ being arbitrary functions of $k$. Therefore, it is easily seen
this solution can be `gauged away' by appropriate choice of the variables 
$\zeta_\mu (k)$ in (\ref{50}) as this corresponds to pure gauge. Thus, analogous to Maxwell theory, the 
massive excitations of linearized Einstein gravity  are gauge artefacts.
 
For massless ($k^2 = 0$) excitations, the equation of motion (\ref{51}) reduces
 to 
\begin{equation}
- k^\mu k_\alpha \varepsilon^{\alpha \nu} -
k^\nu k_\alpha \varepsilon^{\alpha \mu }
+  k^\mu k^\nu \varepsilon
+\eta^{\mu\nu}k_\alpha k_\beta \varepsilon^{\alpha \beta} = 0.
\label{51+3}
\end{equation}
In  a frame  of reference where $k^{\mu} = (\omega , 0, 0, \omega )^T$  
the above equation can be written as 
\begin{equation}
-\omega [k^{\mu} (\varepsilon^{0\nu} - \varepsilon^{3 \nu}) + 
k^{\nu} (\varepsilon^{0\mu} - \varepsilon^{3 \mu})] + k^{\mu} k^{\nu}  \varepsilon +\omega^2 \eta^{\mu\nu}(\varepsilon^{00} + \varepsilon^{33} - 2 \varepsilon^{03}) = 0.
\label{51+4}
\end{equation}
Various components of the above equation together with the symmetricity of 
$\varepsilon^{\mu\nu}$ leads to the reduction in the number of independent
components of $\varepsilon^{\mu\nu}$. In (\ref{51+4}), ${\mu = \nu = 0}$ leads 
to $ \varepsilon^{11} = -\varepsilon^{22}$; ${\mu =1, \nu = 0}$ to 
$ \varepsilon^{01} = -\varepsilon^{31}$;  ${\mu =2, \nu = 0}$ to $ \varepsilon^{02} = -\varepsilon^{32}$;  ${\mu =2, \nu = 0}$ to $ \varepsilon^{03}
 = -\varepsilon^{33}$;  and ${\mu =1, \nu = 1}$ to $ \varepsilon^{00} 
 = -\varepsilon^{03}$. Therefore, one can write the polarization tensor $\varepsilon^{\mu\nu}$ in a reduced form as
\begin{equation}
\{\varepsilon_{\mu\nu}\} = \left(
\begin{array}{cccc}
 \varepsilon^{00} & \varepsilon^{01} &  \varepsilon^{02}& \varepsilon^{00} \\
\varepsilon^{01} & \varepsilon^{11} &  \varepsilon^{12} & \varepsilon^{01} \\
 \varepsilon^{02} & \varepsilon^{12} & -\varepsilon^{11} &  \varepsilon^{02} \\
 \varepsilon^{00} & \varepsilon^{01} &  \varepsilon^{02}& \varepsilon^{00}
\end{array}
\right).
\label{51+5}
\end{equation}
Now with the choice $ \zeta^0 = \zeta^3 = -\frac{\varepsilon^{00}}{\omega}, 
\zeta^2  =-\frac{\varepsilon^{01}}{\omega}, \zeta^2  =- 
\frac{\varepsilon^{02}}{\omega}$, the polarization tensor $\{\varepsilon_{\mu\nu}\}$ can be written in the maximally reduced form (similar to (\ref{9}) of 
Maxwell theory) as follows: 
\begin{equation}
{\cal E} \equiv  \{\varepsilon_{\mu\nu}\} = \left(
\begin{array}{cccc}
0 & 0 & 0 & 0\\
0 & a & b & 0 \\
0 & b & -a & 0 \\
0 & 0 & 0 & 0
\end{array}
\right)
\label{53}
\end{equation}
(which is gauge equivalent to (\ref{51+5})) where $a$ and $b$ are arbitrary parameters representing the two degrees of freedom
for the theory. (This form of the polarization tensor of linearized Einstein 
gravity in 3+1 dimensions is derived in \cite{weinberg1} following a different approach.) 
In this form (\ref{53}),  the polarization tensor ${\cal E}$ satisfies the harmonic gauge condition, 
\begin{equation}
k_\mu \varepsilon^{\mu}_{\nu} = \frac{1}{2} k_\nu \varepsilon^{\mu}_{\mu}
\label{52+0}
\end{equation}in momentum space, 
automatically.
Using the maximally reduced form (\ref{53}) of the polarization tensor, it 
is now
straightforward to show that the group $W(p, q) $ in (\ref{3}) generate
gauge transformations in linearized Einstein gravity also. For this purpose
consider the action of $W(p, q)$ on ${\cal E}$ in (\ref{53}), just as in 
KR case, 
$${\cal E} \rightarrow {\cal E}'= W(p, q) {\cal E} W^T(p, q)$$
\begin{equation}
 = {\cal E} +\left(
\begin{array}{cccc}
((p^2 - q^2)a + 2pqb)  & (pa + qb) & (pb -qa) & ((p^2 - q^2)a + 2pqb) \\
(pa + qb) &0 & 0 & (pa + qb) \\
(pb -qa)  & 0 & 0 & (pb -qa) \\
((p^2 - q^2)a + 2pqb) &( pa + qb) &( pb -qa) & ((p^2 - q^2)a + 2pqb)
\end{array}\right).
\label{54}
\end{equation}
The above transformation can be cast in the form of a gauge transformation (\ref{50}) with the 
following choice for  the arbitrary functions $\zeta^\mu(k)$:
\begin{equation} 
\zeta^0 = \zeta^3 = \frac{(p^2 - q^2)a + 2pqb}{\omega},
\label{55}
\end{equation}
\begin{equation}
\zeta^1 = \frac{pa + qb}{\omega},
\label{56}
\end{equation}
\begin{equation}
\zeta^2 = \frac{pb -qa}{\omega}.
\label{57}
\end{equation}
We therefore conclude that the translational subgroup $T(2)$ of the Wigner's
little group $E(2)$, in its original representation $W(p, q)$ (\ref{3}),     
acts as a  gauge generator in linearized gravity in 3+1 dimensions. 

It is important to notice that, unlike vector gauge (Maxwell) fields, 
the translational subgroup of Wigner's little
group acts as gauge generator for linearized gravity only when the spacetime
has the dimension $d=4$ and not in any other dimension. This can be seen
immediately from the following simple reasoning. In $d$ dimensions,
the translational subgroup of the little group (for massless particles) has  
$d-2$ generators which, incidentally, is the number of (transverse)
 degrees of freedom in
Maxwell theory. On the other hand, the number of degrees of freedom for 
linearized gravity in $d$ dimensions is $\frac{1}{2}d(d-3)$. Therefore,
the number of generators of the translational subgroup equals the number
 of degrees of freedom of the theory, if and only if $d=4$. 
  
\subsection{Linearized Einstein-Chern-Simons(ECS) theory in 2+1 dimensions} 

The full action of the topologically massive gravity in 2+1 dimensions is 
\begin{equation}
I^{ECS} = I^E + I^{CS}
\label{57+1}
\end{equation} 
where the 2+1 dimensional Einstein action here is 
\begin{equation} 
I^E = \int d^3 x \sqrt{g} R 
\label{57+2}
\end{equation}
and the Chern-Simons action $I^{CS}$ is given by 
\begin{equation}
I^{CS} = -\frac{1}{4\mu} \int d^3 x \epsilon^{def}[R_{deab} {\omega_f}^{ab} +
\frac{2}{3}{\omega_{db}}^c {\omega_{ec}}^a{\omega_{fa}}^b].
\label{57+3}
\end{equation}
Here the indices $a,b,c$ etc. stand for local Lorentz indices  in 2+1 dimensions. The   $\omega_{dab}$ are the components of the spin connection one-form and 
are related to the curvature two-form by the 
 second Cartan's equation of structure (${R^a}_b = d{\omega^a}_b  + 
{\omega^a}_c \wedge {\omega^c}_b $). Note that the sign in front of the Einstein action is now opposite to 
the conventional one (\ref{43}) and is required to make the full theory free of ghosts \cite{djt}. 
The linearization, $g_{ab} = \eta_{ab} + h_{ab}$, of the ECS theory (\ref{57+1}) results in the 
abelian theory given by  
\begin{equation}
I^{ECS}_L =\int d^3 x {\cal L}_L^{ECS} 
\label{57+4}
\end{equation}
where
\begin{equation} 
{\cal L}_L^{ECS} = {\cal L}_L^E + {\cal L}_L^{CS}. 
\label{58}
\end{equation} 
Here
\begin{equation}
{\cal L}_L^E = -\frac{1}{2}h_{ab} \left[ R^{ab}_L - \frac{1}{2} \eta^{ab} R_L\right]
\label{59}
\end{equation}
now is the Lagrangian for linearized version of pure gravity in 2+1 dimensions 
and is the same as (\ref{44}) except that it has the opposite sign and the indices, in this case, vary over $0, 1, 2$. Similarly,
\begin{equation}
{\cal L}_L^{CS} = -\frac{1}{2\mu} \epsilon_{abc} 
\left[ R^{bd}_L - \frac{1}{2} \eta^{bd} R_L\right]
\partial^a h^c_d
\label{60}
\end{equation}
is the linearized Chern-Simons term with the  Chern-Simons parameter  $\mu$.
Under the gauge transformation
$h_{ab} \rightarrow h'_{ab} = h_{ab} + \partial_a \xi_b (x) + \partial_b \xi_a (x)$, the
Chern-Simons part ${\cal L}_L^{CS}$ changes by a total derivative:
\begin{equation}
\delta {\cal L}^{CS}_L = \frac{1}{\mu}\epsilon_{abc} \partial_d \left(
R^{bd}_L \partial^a \xi^c\right)
\label{61}
\end{equation}
The equation of motion corresponding to ${\cal L}_L^{ECS}$ is given by
$$- \Box h^{ab} + \partial^a \partial_c h^{cb} + \partial^b \partial_c h^{ca} 
- \partial^a\partial^b h + \eta^{ab} (\Box h - \partial_c\partial_d h^{cd})
$$
\begin{equation}
-\frac{1}{2\mu} \epsilon^{acd} \partial_c (\Box h^b_d 
- \partial_e \partial^b h^e_d) - 
\frac{1}{2\mu} \epsilon^{bcd} \partial_c (\Box h^a_d                        
- \partial_e \partial^a h^e_d) 
= 0.
\label{62}
\end{equation}
With the ansatz  $h_{ab} = \chi_{ab} (k) e^{ik.x} $,  
 the above equation of motion can be written
in terms of the symmetric polarization tensor $\chi_{ab}(k)$ and the 3-momentum
$k^a$ as follows
$$
k^2\chi^{ab} - k^a k_c \chi^{cb} -
k^b k_c \chi^{ca } +  k^a k^b \chi
+\eta^{ab}(-k^2 \chi +k_c k_d \chi^{cd})
$$
\begin{equation}
 -\frac{i}{2\mu}\left[ \epsilon^{acd} k_c (-k^2 \chi^b_d +
k_e k^b \chi^e_d)~ +~ \epsilon^{bcd} 
k_c (-k^2   \chi^a_d
+ k_e k^a \chi^e_d)\right] = 0.
\label{63}
\end{equation}
Analogous to (\ref{50}), the expression for the gauge transformation for
ECS theory in terms of its polarization tensors $\chi_{ab}(k)$ is given by
\begin{equation} 
\chi_{ab}(k) \rightarrow \chi_{ab}'(k) = \chi_{ab}(k) + k_a \xi_b(k) + k_b \xi_a(k)
\label{new63}
\end{equation}
where $\xi_a(k)$ are small arbitrary functions of $k$.  
Depending on whether the excitations are  massless or massive, we have  two
options for $k^2$:\\
\begin{center} (i) $k^2 = 0$ or (ii) $k^2 \neq 0$. \\ \end{center}
{\it case} (i): $k^2 = 0$ \\
Contracting the (\ref{63}) with $\eta_{ab}$ gives 
\begin{equation}
k_a k_b \chi^{ab} = 0.
\label{64}
\end{equation}
A general solution to this equation consistent with the equation of motion (\ref{63}) is 
\begin{equation}
\chi^{ab} = k^a f^b(k) + k^b f^a(k)
\label{65}
\end{equation}
where $f^a(k)$ are arbitrary functions of $k$. However, with $\xi^a = -f^a$ we can
`gauge away' these solutions. Therefore, massless excitations of ECS theory
are pure gauge artefacts.  
We now proceed to the other option:\\
{\it case} (ii) $k^2 \neq 0$. \\
Let $k^2 = m^2$. On contraction with $\eta_{ab}$ (\ref{63}) gives
\begin{equation}
k_a k_b \chi^{a b} = m^2\chi
\label{66}
\end{equation}
where $\chi = {\chi^a}_a$.
With $k^{\alpha} = (m, 0, 0)^T$, this yields
\begin{equation}
\chi_{11} + \chi_{22} = 0.
\label{67}
\end{equation}
By considering the spatial part of (\ref{63} ) one can show that the mass $m$ of the excitations can be identified with the 
Chern-Simons parameter $\mu$ as follows. The spatial part of (\ref{63} )
is 
$$
m^2\chi^{ij} - k^i k_a \chi^{a j} -
k^j k_a \chi^{a i } +  k^i k^j \chi
+\eta^{ij}(-k^2 \chi +k_a k_b \chi^{a b})
$$
\begin{equation}
 -\frac{i}{2\mu}\left[ \epsilon^{i a b} k_a (-k^2 \chi^j_b +
k_e k^j \chi^e_b)~ +~ \epsilon^{j a b}
k_a (-k^2   \chi^i_b
+ k_e k^i \chi^e_b)\right] = 0.
\label{68}
\end{equation}
In this equation $i, j$ takes values $1$ and  $2$.
On passing to the rest frame the above equation simplifies to
\begin{equation}
\chi^{ij} -\chi^{kk}  -\frac{im}{2\mu} \left[
\epsilon^{ik}\chi^j_k + \epsilon^{jk}\chi^i_k\right] = 0
\label{69}
\end{equation}
from which we obtain (for  $i=j=1$ and $i=j=2$  respectively) 
\begin{equation}
\chi^{11} = -\frac{im}{\mu}\chi^{12}~; ~~~~~\chi^{22}=
\frac{im}{\mu} \chi^{12}. 
\label{70}
\end{equation}
With $i=1$ and $j=2$ we have
\begin{equation}
\chi^{12} = -\frac{im}{\mu}\chi^{22}.
\label{71}
\end{equation}
This relation together with (\ref{70}) implies that $m^2 = \mu^2$.
The remaining components can be made to vanish by a suitable gauge choice.
 Finally, for the Chern-Simons parameter $\mu > 0 $ , the polarization tensor 
{\bf $\chi_+$} of the gravity coupled to Chern-Simons
theory in 2+1 dimensions  in the rest frame can be written as 
\begin{equation}
{\bf \chi_+ } = \{\chi_+^{ab}\}= \left(
\begin{array}{ccc}
0 & 0 & 0   \\
0 & 1 & i \\
0 & i & -1 
\end{array}
\right)\tau
\label{72}
\end{equation}
where $\tau$ is an arbitrary real parameter. Notice that the ECS theory 
has only a single degree 
of freedom 
corresponding to the  parameter $\tau$. Similarly, the rest frame
 polarization tensor for
an ECS theory having the Chern-Simons parameter $\mu < 0$ is
\begin{equation}
{\bf \chi_-} = \{\chi_-^{ab}\}= \left(
\begin{array}{ccc}
0 & 0 & 0   \\
0 & 1 & -i \\
0 & -i & -1
\end{array}
\right)\tau.
\label{73}
\end{equation}  

It is important to note that these rest frame polarization tensors {\bf $\chi_\pm$}
of
ECS theories (with $ \tau = \frac{1}{2}$) can be obtained as
direct products of the rest frame polarization vectors $\bar{\varepsilon}_\pm^a$ (\ref{31})
of MCS theories. i.e.,
\begin{equation}
\chi^{ab}_\pm  = \bar{\varepsilon}^{a}_\pm \bar{\varepsilon}^{b}_\pm. 
\label{74}
\end{equation}
This suggests that we adopt   orthonormality conditions for {\bf $\chi_\pm$} which will are similar to
(\ref{36}) that are used for $\bar{\varepsilon}^{a}_\pm$. Hence we require
\begin{equation}
tr\left( (\chi_+)^{\dagger}(\chi_-)\right) = 0~; ~~~~~    tr\left( (\chi_\pm)^{\dagger}(\chi_\pm)\right) = 1.
\label{75}
\end{equation}
Therefore, analogous to (\ref{31}) of MCS theory,  we have the following 
maximally reduced form for the polarization 
tensors of a pair of  ECS theories with opposite helicities  
\begin{equation} 
\chi_\pm = \frac{1}{2}\left(
\begin{array}{ccc}
0 & 0 & 0   \\
0 & 1 & \pm i \\
0 & \pm i & -1
\end{array}
\right).
\label{76}
\end{equation}
Note that these polarization matrices of ECS theories are traceless and 
singular. 
We are now equipped to study the role
played by the
translational group in
generating the gauge transformation in this theory. The representations of
$T(1)$ that generates gauge transformation in pair of MCS theories with
opposite helicities is given by $G_\pm (\alpha)$ (\ref{41}). On account of the
relation (\ref{74}) it is expected that
the same representation will generate gauge transformations in ECS theories
also. Indeed one can easily see that $G_\pm(\tau_\pm) $ are the gauge generators in
ECS theories:
\begin{equation}
\chi_\pm \rightarrow \chi_\pm '=  G_\pm (\tau_\pm) \chi_\pm  G^T_\pm (\tau_\pm ) = \chi_\pm
+  \left(\begin{array}{ccc}
\tau_\pm^2 & \frac{\tau_\pm}{\sqrt{2}} & \frac{\pm i\tau_\pm}{\sqrt{2}}  \\
 \frac{\tau_\pm}{\sqrt{2}} & 0 & 0 \\
 \frac{\pm i\tau_\pm}{\sqrt{2}} & 0 & 0
\end{array}
\right).
\label{79+0}
\end{equation}
This transformation  can be cast  in the form of the  gauge transformation 
(\ref{new63}) with the following choice of $\xi$'s;
\begin{equation} 
\xi_0 = \frac{\tau_\pm^2}{{2}|\mu |}, ~~~ \xi_1 = \frac{\tau_\pm}{\sqrt{2}|\mu |},~~~ \xi_2 = \frac{\pm i\tau_\pm}{\sqrt{2} |\mu |}.\label{79+1}
\end{equation}  
One can obtain the moving frame expression for  polarization tensors 
{\bf $\chi_\pm(k)$} from the above 
rest frame results by applying appropriate Lorentz boost  as follows:
$${\bf \chi_\pm}(k) = \Lambda^T(k) {\bf \chi_\pm}(0)  \Lambda(k)$$
\begin{equation}  
 = \frac{1}{2\mu^2}\left(\begin{array}{ccc}
k^2_0 - \mu^2 & k^0 k^1 \mp i\mu k^2 &  k^0 k^2 \pm i\mu k^1 \\
k^0 k^1 \mp i\mu k^2 & \frac{(k^0 k^1 \mp i\mu k^2)^2}{k^2 _0 - \mu^2} &
\frac{(k^0 k^1 \mp i\mu k^2)( k^0 k^2 \pm i\mu k^1)}{k^2 _0 - \mu^2}  \\
k^0 k^2 \pm i\mu k^1 & \frac{(k^0 k^1 \mp i\mu k^2)( k^0 k^2 \pm i\mu k^1)}{k^2 _0 - \mu^2} & \frac{(k^0 k^2 \pm i\mu k^1)^2}{( k^0 k^2 \pm i\mu k^1)^2}
\end{array}
\right) e^{\pm 2i \phi(k)}
\label{77}
\end{equation} 
where $ \phi(k) = \arctan (\frac{k^2}{k^1})$ and  the momentum space boost matrix
\begin{equation}
\Lambda (k)~~ =~~ \left( \begin{array}{ccc}
   {\gamma} & {\gamma}{\beta}^1 & {\gamma}{\beta}^2      \\
        {\gamma}{\beta}^1 & 1 + \frac{({\gamma} -1)({\beta}^1)^2}{(\vec{\beta})^2} & \frac{({\gamma} -1){\beta}^1{\beta}^2}{(\vec{\beta})^2} \\
         {\gamma}{\beta}^2 & \frac{({\gamma} -1){\beta}^1{\beta}^2}{(\vec{\beta})^2} & 1 + \frac{({\gamma} -1)({\beta}^2)^2}{(\vec{\beta})^2}
    \end{array} \right)
\label{78}
\end{equation} 
with $\vec{\beta} = \frac{\bf k}{k^0}$ and $\gamma = \frac{k^0}{|\mu |}$.
Identical results were obtained in other contexts \cite{grignani} by different
methods. Here, we have shown how these results can be obtained in a simpler and
straightforward manner just by considering the momentum space expression of the 
equation of motion in the rest frame using the plane wave method with
 a subsequent boost transformation.
Obviously, the relation (\ref{74}) holds true in the moving frame also.

\section{Dimensional descent and linearized gravity theories}
In order to discuss dimensional descent from 3+1 dimensions to 2+1 dimensions for
symmetric 2nd rank tensor fields it is essential to discuss the relevant aspects 
of 2+1 dimensional Einstein-Pauli-Fierz(EPF) theory whose action  is given by
\begin{equation}
I^{EPF} = \int d^3 x \left(- \sqrt{g} R - \frac{\mu^2}{4}(h_{ab}^2 - h^2)     
 \right). 
\label{79}
\end{equation}
Note that the usual sign in front of Einstein action has been restored to avoid 
ghosts and tachyons, as has been observed recently by Deser and Tekin 
\cite{tekin}. As noted in \cite{tekin}, both the relative and and overall signs
of the two terms in (\ref{79}) have to be of conventional Einstein and Pauli-Feirz mass terms in order to have a physically meaningful theory\footnote{On the other hand in the ECS theory, sign of the Einstein term has to be 
opposite to that of the conventional one for the theory to be viable. Therefore,  
if one attempts to couple the ECS theory with a Pauli-Fierz term, one is faced
 with an unavoidable conflict of signs.}. Upon linearization, (\ref{79}) reduces to
\begin{equation}
{\cal L}_L^{EPF} = \frac{1}{2}h_{ab} \left[ R^{ab}_L - 
\frac{1}{2} \eta^{ab} R_L\right] - \frac{\mu^2}{2}(h_{ab}^2 - h^2).
\label{80}
\end{equation}
Analogous to the doublet structure of Proca theory discussed above, 
the EPF theory
is a doublet, as was suggested in \cite{djt},  comprising of a pair of ECS theories having opposite helicities. 
And just like the Proca theory,
EPF theory is a gauge noninvariant  theory. The equation of motion following from the EPF Lagrangian is
given by 
\begin{equation} 
- \Box h^{ab} + \partial^a \partial_c  h^{cb} + \partial^b\partial_c  h^{ca}
 - \partial^a \partial^b h + \eta^{ab}(\Box h -\partial_c \partial_d h^{cd}) - \mu^2(h^{ab} - \eta^{ab} h) = 0. 
\label{81}
\end{equation}
With the ansatz $ h^{ab} = {\cal X}^{ab} e^{ik.x}$, where ${\cal X}^{ab}$ is the polarization tensor in this case,  this equation can
be written as
\begin{equation}
k^2{\cal X}^{ab} - k^a k_c {\cal X}^{cb} -
k^b k_c {\cal X}^{ca } +  k^a k^b {\cal X}
+\eta^{ab}(-k^2 {\cal X} +k_c k_d {\cal X}^{cd}) -
\mu^2({\cal X}^{ab} - \eta^{ab}{\cal X}) = 0
\label{82}
\end{equation}
We now proceed along the same lines as was done in the previous cases 
to arrive at the physical polarization tensor of EPF theory. If we  choose $k^2 =0$, the 
equation of motion (\ref{82}) leads, upon contraction with $k_a$, to the 
condition 
\begin{equation}
k_a {\cal X}^{ab} = k^b{\cal X}. \label{821}
\end{equation} 
On the other hand, the  contraction of (\ref{82}) with $\eta_{ab}$ leads to
\begin{equation}
{\cal X} + k_a k_b {\cal X}^{ab} = 0.\label{822}
\end{equation}
A solution of the above pair of  equations (\ref{821}, \ref{822}) is given by
$  {\cal X}^{ab} = k^af^b(k) + k^b f^a(k)$ where $f$'s are arbitrary functions of 
$k$. This solution automatically satisfies the condition $k.f=0$. However,
such a solution is compatible with  the equation of motion (\ref{82}) if and 
only if the $f$'s vanish identically. Thus the EPF theory does not have 
massless excitations. 

Now for $k^2 \neq 0$, in the rest frame $k^{\mu} = (m, 0, 0)^T$ and so the $(00)$ component of the equation of motion yields
$\mu^2({\cal X}^{00} - {\cal X}) = 0$ which in turn gives 
\begin{equation}
{\cal X}^{11} + {\cal X}^{22} = 0.
\label{82+1}
\end{equation}
Therefore, one is free to arbitrarily choose either ${\cal X}^{11}$ or ${\cal X}^{22}$. 
Similarly, the  $(0i)$ component in the rest frame becomes $\mu^2  {\cal X}^{0i} = 0$  implying
\begin{equation}
{\cal X}^{0i} = 0.
\label{82+2}
\end{equation} 
The space part ($(ij)$-components) of the equation of motion with $i=j=1$ 
and $i=j=2$, respectively yields in the rest frame,
\begin{equation}
-\mu^2 ({{\cal X}^0}_0 + {{\cal X}^2}_2) + m^2 {{\cal X}^1}_1  = 0
\end{equation}
\begin{equation}
-\mu^2 ({{\cal X}^0}_0 + {{\cal X}^1}_1) + m^2 {{\cal X}^2}_2  = 0.
\end{equation}
Adding the above to equation gives,
\begin{equation}
{\cal X}^{00} = 0.
\label{82+3}
\end{equation}
Finally, the $(ij)$ compnent of  (\ref{82}) for $i\ne j$ 
in the rest frame becomes
\begin{equation}
(m^2 - \mu^2){{\cal X}^1}_2 = 0.
\end{equation}
which can be satisfied if either $(m^2 - \mu^2) = 0$ (in which case ${{\cal X}^1}_2$ may remain arbitrary) or ${{\cal X}^1}_2 = 0.$ The latter case implies 
that we will have only one degree of freedom in the theory corresponding to
${{\cal X}^1}_1$ (or ${{\cal X}^2}_2$). However, as is well known, the EPF theory
has two degrees of freedom \cite{djt}. Therefore, we must have
\begin{equation}
m^2 = \mu^2
\label{82+4}
\end{equation}
thus establishing that the mass of the EPF excitation to be $|\mu |$ and leaving
 the $\chi^{12}$ component  arbitrary. 
The two massive degrees of freedom of EPF model correspond to ${\cal X}_{11} = 
-{\cal X}_{22} = a$ and $ {\cal X}^{12} =  {\cal X}^{21} = b$ where $a, b$ are arbitrary parameters. 
Therefore,from (\ref{82+2} ,\ref{82+3} , \ref{82+4}) it is obvious that we can write the polarization 
tensor of EPF
 model in the rest frame as
\begin{equation}
{\cal X} =
 \left(
\begin{array}{ccc}
0 & 0 & 0   \\
0 & a & b \\
0 & b & -a
\end{array}
\right).
\label{83}
\end{equation}

With the aid of the expressions (\ref{83}) for ${\cal X}$ of EPF theory  
and (\ref{76}) for  $\chi_\pm$ of a pair of ECS theories, we now embark on a discussion of
dimensional descent for the case of 2nd rank symmetric tensor gauge fields emphasising
the near exact parallel with case of vector gauge fields. 
One can obtain the 
momentum 3-vector $k^a$ and polarization tensor ${\cal X}$ (\ref{83}) of EPF model in 2+1 dimensions
from those of linearized gravity in 3+1 dimensions as follows.  
By applying the projection operator ${\cal P} = $diag$(1, 1, 1, 0)$ 
on momentum 4-vector
 $k^\mu = (\omega, 0, 0, \omega )^T $ of a massless graviton moving in the $z$-direction of 3+1 dimensional linearized Einstein gravity and subsequently
deleting the last row of the resulting vector, one get momentum 3-vector $k^a$ of 2+1 dimensional
EPF quanta at rest. By a similar application of ${\cal P}$ on ${\cal E}$ (\ref{53}) and deleting
the last row and column, one gets the polarization tensor ${\cal X}$ (\ref{83})  in the rest frame of the EPF quanta.
Next we notice that, just like the way Proca polarization vector ${\bar{\varepsilon}}^a$ is written as a linear combination of two orthonormal canonical vectors (\ref{34}), one can write the EPF
polarization tensor ${\cal X}$ as 
\begin{equation}
{\cal X} = a {\cal X}_1 + b {\cal X}_2 =  a \left(
\begin{array}{ccc}
0 & 0 & 0   \\
0 & 1 & 0 \\
0 & 0 &-1
\end{array}
\right) + b \left(\begin{array}{ccc}
0 & 0 & 0   \\
0 & 0 & 1 \\
0 & 1 & 0
\end{array}
\right).
\label{84}
\end{equation}
We may consider the above equation to be the EPF analogue of (\ref{34}) in the case
Proca theory. 
Notice that the space parts of the matrices appearing in the above linear
combination are nothing but the Pauli matrices 
\begin{equation}
\sigma_1 = \left(\begin{array}{cc}
 0 & 1 \\
1 & 0
\end{array}
\right),  ~~~~ \sigma_3 = \left(\begin{array}{cc}
 1 & 0 \\
0 & -1
\end{array}
\right).
\end{equation}
Clearly, the space part $\{{\chi_\pm}^{ij}\}$ of the ECS polarization tensors $\chi_\pm$ (\ref{76}) can be expressed in terms of $ \sigma_1$ and $\sigma_3$ 
as follows:
\begin{equation} 
\{{\chi_\pm}^{ij}\} = \frac{1}{\sqrt{2}}(\frac{1}{\sqrt{2}}\sigma_3 \pm \frac{i}{\sqrt{2}} \sigma_1 )~.
\label{85-1}
\end{equation}
This amounts to an SU(2) transformation in the 2-dimensional subspace (of the
SU(2) Lie algebra in an orthonormal basis) spanned by  
$\frac{1}{\sqrt{2}}\sigma_1$ and $ \frac{1}{\sqrt{2}} \sigma_3$\footnote{Note 
that the $\{\chi^{ij}_-\}$ obtained in (\ref{85-1}) differs from the one 
obtained by SU(2) rotation by an irrelevant $i$ factor just as in the vector case.}. 
It should be noticed that (\ref{85-1}) is the analogue of  
  (\ref{37-1}) in the case of Proca and MCS theories. In the case of vector
  (Proca and MCS) field theories, the basis vectors $\varepsilon_1$ and 
$\varepsilon_2$ (\ref{34}) of the Proca polarization vector, when transformed 
by a suitable SU(2) transformation yield the polarization vectors 
$\varepsilon_\pm$
of a pair of MCS theories. Similarly, in the case of tensor (EPF and ECS) field
 theories, the same SU(2) transformation when acted on the ${\cal X}_1$ and 
${\cal X}_2$ provides the polarization tensors $\chi_\pm$ of a doublet of 
ECS theories with opposite helicities just as  Proca theory is a doublet
of MCS theories having  opposite spins.
This corroborates the proposition that EPF theory is consisted of 
a doublet of ECS theories with opposite spins at least at the level of 
polarization tensor. Moreover, as we have discussed earlier, 
the polarization tensor and momentum vector of (2+1 dimensional) EPF theory can be obtained 
from those of linearized Einstein gravity (in 3+1 dimensions) by applying suitable projection operator. This relationship between EPF and ECS theories resemble the one between Proca and MCS theories. Therefore we expect that 
the procedure of dimensional descent to be valid here as well. As described in section 2, 
the generator of the  representations of $T(1)$, obtained by dimensional descent,  that generate 
gauge transformation in a pair of MCS theories with opposite helicities are 
given by
$\bar{T}_\pm$  (\ref{37}). 
Also, the ECS polarization tensors $\chi_\pm$ can be made to 
satisfy the orthonormality relations
(\ref{75}) similar to (\ref{36}) for MCS case owing to the fact that the 
former is a tensor product of MCS polarization  vectors. 
Hence it is natural to expect that the $T(1)$ group representation $G_\pm (\tau)$ (\ref{41}) obtained by exponentiation of $\bar{T}_\pm$ generates gauge 
transformation in  ECS doublet, which in fact it does, as we have shown in
(\ref{79+0}).
Therefore, it is evident that by a dimensional descent from 3+1 dimensional linearized gravity
one could obtain the representations of $T(1)$ that generate gauge transformations in the doublet of
topologically massive ECS theories in 2+1 dimensions. This is similar to  the dimensional descent
from 3+1 dimensional Maxwell theory to 2+1 dimensional MCS theory discussed in section 2. 

\section{Conclusion}
 
In this paper we have studied the role  of translational subgroup of
Wigner's little group for massless particles, as a generator of gauge
transformations in 3+1 dimensional linearized (pure) Einstein gravity and
2+1 dimensional ECS theory which are  2nd rank symmetric tensor field theories.
This
property of translational group was earlier shown to hold for vector
gauge theories and 2-form gauge fields in different dimensions. In those cases
the theories considered included ordinary  gauge fields as well as topologically
massive  gauge fields. In section 2, a review of the role of translational group  in generating gauge transformations in the vector gauge 
( Maxwell and MCS) theories  and in 2nd rank antisymmetric tensor gauge
(KR and $ B\wedge F$) theories is provided. Taking the example of
3+1 dimensional Maxwell and 2+1 dimensional Proca theory (which is a doublet of
MCS theories with opposite helicities), we have  illustrated the method of
dimensional descent by which one obtains the representation of the little
group that generates gauge transformation in a topologically massive theory.
In section 3, we studied the linearized Einstein gravity in 3+1 dimensions
and by a straightforward analysis of its the equation of motion, using the plane wave method in momentum
space, we obtatined the maximally reduced form of the polarization tensor
of the theory and the result agrees with the expression of the polarization 
tensor obtained by other
methods \cite{weinberg1}.  
Using this maximally reduced polarization tensor, we have shown
explicitly that $T(2)$, the translational subgroup  of the the Wigner's little
group $E(2)$, generates gauge transformations in 3+1 dimensional linearized
gravity just as it happens in the cases of  vector or Kalb-Ramond theory
involving  antisymmetric tensor
gauge fields. In the next section we consider the topologically massive ECS 
theory, obtained
by coupling a Chern-Simons mass term to the linearized gravity in 2+1 dimensions and show that the polarization tensor of ECS theory is actually  a tensor
product of the polarization vectors of a pair of MCS theories with the same
Chern-Simons parameter. We have further shown that the same representation of
$T(1)$ that generate gauge transformation in MCS theories acts as generator
of gauge transformation in ECS theory also. Finally, in section 4 we
obtain the polarization tensor of EPF theory 
(obtained by coupling the linearized
gravity to a Pauli-Fierz mass term) in 2+1 dimensions and show that it splits
into  ECS polarization tensors with opposite helicitites thus suggesting a
doublet stucture for EPF theory at the level of polarization tensors. 
This is very similar to Proca theory which
is a doublet of MCS theories. Drawing this analogy further, we have been
able to extend the method of dimensional descent to the case of second
rank symmetric tensor field theories and show that one can obtain the EPF polarization
tensor from that of linearized Einstein gravity in 3+1 dimensions.
 Further more, one obtains the   representation of $T(1)$ that generate
gauge transformation in ECS theory by following   exactly the same procedure
as that  in the case of MCS theory.

However, it should be mentioned that the the analogy between EPF and Proca 
theories with their respective doublet structures breaks down if one considers
the fact that sign of the Einstein term flips from EPF to ECS theories in
contrast to Proca theory where the sign of the Maxwell term remains unchanged
irrespective of whether it is coupled to a Chern-Simons term or a usual mass 
term.
Therefore, further investigations are necessary in order to rigorously 
establish the doublet structure, if any, of EPF theory beyond the level of polarization
tensors.

{\large \bf Acknowledgment:} Authors wish to thank Dr. R. Banerjee for many 
illuminating and useful discussions.

\end{document}